\documentclass[pra,floatfix,twocolumn,superscriptaddress]{revtex4}

\usepackage{graphicx}
\usepackage{amsmath}
\usepackage{bm}

\def\del{\nabla}

\def\bE{\mathbf{E}}\def\bEe{\mathbf{E}_e}\def\bEp{\mathbf{E}_p}
\def\bAp{\mathbf{A}_p}\def\bbAp{\bar{\mathbf{A}}_p}
\def\bAe{\mathbf{A}_e}
\def\bP{\mathbf{P}}\def\bPNL{\mathbf{P}_{\mathrm{NL}}}
\def\bPel{\mathbf{P}_{\mathrm{el}}}

\def\bProt{\mathbf{P}_{\mathrm{rot}}}

\def\bPpl{\mathbf{P}_{\mathrm{pl}}}

\def\br{\mathbf{r}}
\def\chit{\chi^{(3)}}

\def\bxhat{\hat{\mathbf{x}}}\def\byhat{\hat{\mathbf{y}}}\def\bzhat{\hat{\mathbf{z}}}
\def\bke{\mathbf{k}_e}\def\bkp{\mathbf{k}_p}
\def\bk{\mathbf{k}}
 
\def\etal{\textit{et al.~}}

\begin{document}

\title{Effect of two-beam coupling in strong-field optical pump-probe experiments}
\author{J. K. Wahlstrand}
\affiliation{Institute for Research in Electronics and Applied Physics, University of Maryland, College Park, MD 20742, USA}
\author{J. H. Odhner}
\author{E. T. McCole}
\affiliation{Department of Chemistry, Temple University, Philadelphia, PA 19122, USA}
\affiliation{Center for Advanced Photonics Research, Temple University, Philadelphia, PA 19122, USA}
\author{Y.-H. Cheng}
\author{J. P. Palastro}
\affiliation{Institute for Research in Electronics and Applied Physics, University of Maryland, College Park, MD 20742, USA}
\author{R. J. Levis}
\affiliation{Department of Chemistry, Temple University, Philadelphia, PA 19122, USA}
\affiliation{Center for Advanced Photonics Research, Temple University, Philadelphia, PA 19122, USA}
\author{H. M. Milchberg}
\affiliation{Institute for Research in Electronics and Applied Physics, University of Maryland, College Park, MD 20742, USA}

\email{wahlstrj@umd.edu}

\begin{abstract}
Nonlinear optics experiments measuring phase shifts induced in a weak probe pulse by a strong pump pulse must account for coherent effects that only occur when the pump and probe pulses are temporally overlapped.
It is well known that a weak probe beam experiences a greater phase shift from a strong pump beam than the pump beam induces on itself.
The physical mechanism behind the enhanced phase shift is diffraction of pump light into the probe direction by a nonlinear refractive index grating produced by interference between the two beams.
For an instantaneous third-order response, the effect of the grating is to simply double the probe phase shift, but when delayed nonlinearities are considered, the effect is more complex.
A comprehensive treatment is given for both degenerate and nondegenerate pump-probe experiments in noble and diatomic gases.
Results of numerical calculations are compared to a recent transient birefringence measurement [Loriot \emph{et al.}, Opt. Express \textbf{17}, 13429 (2009)] and a recent spectral interferometry experiment [Wahlstrand \emph{et al.}, Phys. Rev. A \textbf{85}, 043820 (2012)].
We also present results from two new experiments using spectrally-resolved transient birefringence with 800 nm pulses in Ar and air and degenerate chirped pulse spectral interferometry in Ar.
Both experiments support the interpretation of the negative birefringence at high intensity as arising from a plasma grating.
\end{abstract}

\maketitle

\section{Introduction}
The intensity-dependent refractive index due to odd-order optical nonlinearities is a fundamental phenomenon in nonlinear optics. Accurate knowledge of it is important for most applications of intense laser pulses.
In condensed matter, the Z scan technique, a measurement of self-focusing \cite{sheik-bahae_high-sensitivity_1989}, is widely employed, but in gases, it is difficult to use because the interaction length cannot be easily characterized.
Experiments that measure the phase shift induced by a pump pulse on itself in an extended medium [2-5] %\cite{nibbering_determination_1997,liu_direct_2005,pscan,laban_self-focusing_2010}
also require careful consideration of propagation effects and plasma defocusing at high intensity \cite{whalen_self-focusing_2011}.
Accounting for these phenomena quantitatively requires extensive theoretical modeling.
For this reason, pump-probe experiments, which measure the effect of cross phase modulation, have an important advantage. 
For a weak probe one can at least be certain that the response of the medium is linear in the probe field, and interpretation of the experimental result appears relatively straightforward.
However, in such an experiment, a pump-probe interference grating \emph{always} appears in the medium and pump light is diffracted by this grating into the probe beam direction.
One must properly account for the contribution of this to the probe phase shift.

Depending on the relative phase between the diffracted pump beam light and the probe beam, the probe can experience a phase shift and/or an amplitude change \cite{smolorz_femtosecond_2000}.
This is the physical mechanism responsible for the enhanced phase shift in cross phase modulation compared to self-phase modulation.
The amplitude effect is often referred to as \emph{two-beam coupling} \cite{boyd}.
For an instantaneous optical Kerr nonlinearity, one finds that the probe picks up twice the nonlinear phase shift of the pump, but there is no net energy transfer.
In general the result depends on the details of the nonlinearity, such as the response time, and even the details of the pulse shape, such as whether or not it is chirped.
A major goal of this paper is to provide a comprehensive discussion of the \emph{phase shift} imparted on the probe pulse and the underlying physics.

We approach this topic in the context of a debate over the higher-order Kerr effect, ignited by a recent degenerate pump-probe experiment in Ar, N$_2$, and O$_2$ reported by Loriot \emph{et al.}~\cite{loriot_measurement_2009}.
A negative birefringence was observed at high pump intensity when the pump and probe pulses overlapped in time.
It was concluded that the expansion of the nonlinear refractive index in powers of the optical intensity $I$, $\Delta n = n_2 I + n_4 I^2 + n_6 I^3+...$, is dominated by higher-order negative terms at high intensity.
Initially, the enhancement of the phase shift in cross phase modulation was not considered, and the Kerr coefficients for the effect of a pulse on itself were corrected in an erratum \cite{loriot_measurement_2010}.
However, this correction accounts only for the nearly instantaneous bound electronic component of the nonlinearity.
An alternative explanation for the negative birefringence observed at high intensity by Loriot \emph{et al}.~\cite{loriot_measurement_2009} can be found by considering the diffraction effects caused by the plasma grating generated when the pump intensity is high enough to significantly ionize the gas \cite{wahlstrand_effect_2011}.
Recently an experiment in air using 400 nm pulses \cite{odhner_ionization-grating-induced_2012} found that the intensity dependence of the negative birefringence was consistent with a calculation of the plasma grating-induced birefringence based on multiphoton ionization \cite{wahlstrand_effect_2011}.

Here, we generalize the theory to allow the use of any ionization model, account for grating effects from the molecular alignment component of the nonlinearity, and extend the theory to handle pulses of arbitrary frequency and chirp.
We also present new experimental results from degenerate pump-probe experiments demonstrating the probe phase shift originating from plasma grating diffraction effects.
We note that many recent experiments have used a nondegenerate probe \cite{chen_measurement_2007,chen_single-shot_2007,feng_direct_2011,wahlstrand_optical_2011,wahlstrand_absolute_2012,odhner_ionization-grating-induced_2012,wahlstrand_high_2012},
for which there is no plasma or rotational grating contribution to the nonlinear phase shift \cite{wahlstrand_effect_2011,odhner_ionization-grating-induced_2012}, as detailed below. 
We first describe a theoretical framework for the calculation of the outgoing probe field in a general pump-probe experiment.
We then numerically calculate the signal and compare the calculations to the Loriot \emph{et al.}~experiment \cite{loriot_measurement_2009}, and to nondegenerate spectral interferometry measurements \cite{chen_measurement_2007,chen_single-shot_2007,wahlstrand_optical_2011,wahlstrand_absolute_2012,wahlstrand_high_2012}.
Finally, we present new experimental results using spectrally-resolved transient birefringence and spectral interferometry with a degenerate probe pulse.

\section{Theory}
Initially reported, as far as we are aware, by Chiao, Kelley, and Garmire \cite{chiao_stimulated_1966}, the grating-induced probe phase shift (``weak wave retardation'') is related to ``self-diffraction'' effects that have been often encountered in experiments in condensed matter (for example, \cite{von_jena_coherent_1979,kalt_laser-induced_1985,yan_influence_2008,tong_femtosecond_2012}).
Two-beam coupling, which typically refers to energy exchange between the pump and probe beams, has been studied in depth \cite{yeh_two-wave_1989,smolorz_femtosecond_2000,dogariu_purely_1997,tang_time-domain_1997,sylla_degenerate_2001,liu_2010}.
However, the probe \emph{phase shift} does not seem to have been discussed except for instantaneous ($n_2$) and photorefractive nonlinearities \cite{yeh_two-wave_1989,yeh_exact_1986}.

Our general approach is similar to previous calculations of two-beam energy coupling \cite{smolorz_femtosecond_2000}.
We assume two incident optical fields, a pump (``excitation'') pulse
\begin{math}
\bEe (\br, t) = (1/2) \bAe (\br, t) e^{i \bke \cdot \br-i\omega_e t} + c.c.,
\end{math}
and a probe pulse
\begin{math}
\bEp^{\mathrm{inc}} (\br, t) = (1/2) \bAp^{\mathrm{inc}} (\br, t) e^{i \bkp \cdot \br-i\omega_p t} + c.c. 
\end{math}
In the experiments in gases that are the primary focus of this paper, the lowest optical resonances are at roughly 10 eV, so the linear response of the medium is nearly instantaneous, and the effect of dispersion over a typical interaction length of a few mm is negligible.
We include the linear response of the neutral, isotropic gas in a constant real refractive index $n_0$ and group the rest of the response in a nonlinear polarization $\bPNL$.
We solve the wave equation
\begin{equation}
\del^2 \bE - \frac{n_0^2}{c^2} \frac{\partial^2\bE}{\partial t^2} = \frac{4\pi}{c^2} \frac{\partial^2\bPNL}{\partial t^2},
\label{wave1}
\end{equation}
in the interaction region, with the incident fields as an initial condition.

In experiments, the probe beam is isolated from the pump beam spatially \cite{loriot_measurement_2009,odhner_ionization-grating-induced_2012} and/or by frequency filtering [16-20]. %\cite{chen_measurement_2007,chen_single-shot_2007,feng_direct_2011,wahlstrand_optical_2011,wahlstrand_absolute_2012}.
In the calculation we are therefore interested only in terms that propagate in the direction $\bkp$ and oscillate at frequencies near $\omega_p$.
This outgoing probe field is
\begin{math}
\bEp (\br, t) = (1/2) \bAp(\br,t) e^{i \bkp \cdot \br-i\omega_p t} + c.c.
\end{math}
We define $\bkp = k_p \bzhat$, $\bke = k_e (\bzhat \cos \theta  + \byhat \sin \theta)$, and linearly polarized incident pump and probe envelopes $\bAe = \bxhat A_e$ and $\bAp^{\mathrm{inc}} = \bxhat A_x^{\mathrm{inc}} + \byhat A_y^{\mathrm{inc}}$.
The outgoing probe envelope, which is in general not linearly polarized, is decomposed according to $\bAp = \bxhat A_x + \byhat A_y$.
To allow for chirped pulses to be handled, $\bAe$ and $\bAp^{\mathrm{inc}}$ are complex.
We neglect diffraction and nonlinear propagation effects on the pump pulse.
The primary pump depletion mechanisms are ionization and pumping of rotational states, which are negligible for the intensities and interaction lengths used in the experiments to which we compare our calculations.

The interaction between pump and probe beams occurs near the waists of nearly Gaussian beam profiles, and in practical experimental implementations, the beam waist is much larger than the wavelength of the light.
The transverse derivatives $\partial^2/\partial x^2$ and $\partial^2/\partial y^2$ in $\del^2$ are thus negligible in the interaction region, and the left side of Eq.~(\ref{wave1}), keeping only terms propagating in the probe direction, becomes (omitting the complex conjugate)
\begin{multline*}
\frac{e^{ik_p z-i\omega_p t}}{2} \left[ -k_p^2 \bAp +2ik_p \frac{\partial \bAp}{\partial z} +\frac{\partial^2 \bAp}{\partial z^2} \right. \\ \left. -\frac{n_0^2}{c^2} \left( -\omega_p^2 \bAp-2i\omega_p \frac{\partial \bAp}{\partial t} + \frac{\partial^2 \bAp}{\partial t^2} \right) \right].
\end{multline*}
The pulses simulated are many cycles long (40 fs and above), so the slowly-varying envelope approximation (SVEA) may be employed, using $|\partial^2\bAp/\partial t^2| \ll |\omega_p \partial \bAp/\partial t|$ and $|\partial^2\bAp/\partial z^2| \ll |k_p \partial \bAp/\partial z|$.
Using $k_p = n_0 \omega_p/c$, we have
\begin{eqnarray}
k_p\left( \frac{in_0}{c} \frac{\partial\bAp}{\partial t} + i \frac{\partial\bAp}{\partial z} \right) e^{ik_p z-i\omega_p t} = \frac{4\pi}{c^2} \frac{\partial^2\bP_{\mathrm{NL},p}}{\partial t^2},
\label{wave3}
\end{eqnarray}
where $\bP_{\mathrm{NL},p}$ includes only terms in $\bPNL$ containing $e^{i\bkp \cdot \br-i\omega_p t}$ (as only they significantly affect the probe field).

We next derive expressions for the polarization $\bP_{\mathrm{NL},p}$ due to the nonlinear interaction of the pump and probe pulses with the medium.
We assume $\bPNL = \bPel + \bPpl + \bProt$, where $\bPel$ is the bound electronic nonlinearity, $\bPpl$ is the plasma (ionization/free electron) nonlinearity, and $\bProt$ is the nonlinearity due to the rotational response (molecular alignment).

\subsection{Electronic nonlinearity}
Cross phase modulation of a weak beam due to a third-order nonlinearity is well known and is treated in nonlinear optics textbooks (e.g., \cite{boyd}).
We provide details here in order to make a clear connection to the more complicated nonlinearities arising from ionization and molecular alignment considered later.
Assuming only a third-order contribution to the polarization and a uniform medium, we have
\begin{multline*}
\left( \bPel \right)_i (\br, t) = \iiint \chit_{ijkl} (t-t_1,t-t_2,t-t_3) \\ \times E_j(\br, t_1) E_k(\br, t_2) E_l(\br, t_3) dt_1 dt_2 dt_3,
\end{multline*}
where $\chit$ is the third-order nonlinear response function.
As all electronic resonances are far higher in energy than $\hbar \omega_e$ and $\hbar \omega_p$, to a good approximation $\chit(t-t_1,t-t_2,t-t_3)\approx\chit \delta(t-t_1) \delta(t-t_2) \delta(t-t_3)$.
The nonlinear electronic component of the polarization then simplifies to
\begin{equation}
(\bPel)_i (\br, t) = \chit_{ijkl} E_j(\br, t) E_k(\br, t) E_l(\br, t).
\label{Pchi3}
\end{equation}

We use the total field $\bEe+\bEp$ in Eq.~(\ref{Pchi3}).
We emphasize that only the terms propagating in the $\bkp$ direction, which are proportional to $e^{i\bkp \cdot \br-i\omega_p t}$, contribute to the signal in the experiment.
The terms proportional to $e^{i\bke \cdot \br-i\omega_e t}$ lead to changes in the phase and amplitude of the pump field, which we neglect.
Terms proportional to $e^{i(2\bke-\bkp)\cdot \br-i(\omega_e-\omega_p)t}$, and $e^{i(2\bkp-\bke)\cdot \br-i(\omega_p-\omega_e)t}$, lead to optical fields that do not reach the detector in properly designed pump-probe experiments.
In degenerate experiments the probe beam is isolated spatially, and in nondegenerate experiments, where typically the pump and probe beams are collinear, the probe beam is isolated spectrally.
Terms proportional to $e^{i(3\bke)\cdot \br - i(3\omega_e)t}$ are responsible for third-harmonic generation.
We do not include the effective nonlinear refractive index caused by harmonic generation through cascading \cite{bache}.
For $\sim$0.1 atm Ar, N$_2$, and O$_2$ and typical interaction lengths of a few mm used in experiments, harmonic cascading from $\chi^{(3)}$, which causes a negative effective $n_4$, is negligible in the components of air \cite{bache}.
%In addition, in a degenerate pump-probe experiment, the effective nonlinear refractive index for the probe beam due to cascading depends strongly on the pump-probe crossing angle $\theta$ due to wavevector mismatch, and is negligible for typical values of $\sim 3^\circ$ in degenerate experiments \cite{wahlstrand_unpub}.

In the SVEA, it is shown in Appendix A that, defining the Kerr coefficient $n_2 = 12\pi^2 \chi^{(3)}_{xxxx}/(n_0^2 c)$,
we have for the $x$ component of the nonlinear polarization term,
\begin{multline}
\frac{4\pi}{c^2} \frac{\partial^2(\bPel)_x}{\partial t^2} = \frac{2 n_0 n_2 \omega_p}{c^2}  \left(   - \omega_p I_e A_x  -2i \frac{\partial I_e}{\partial t} A_x \right. \\ \left. -2i I_e \frac{\partial A_x}{\partial t} \right) e^{i\bkp \cdot \br -i\omega_p t},
\label{finalKerr}
\end{multline}
where $I_e = n_0 c/(8\pi) |A_e|^2$ is the pump intensity.
Off-diagonal elements of $\chi^{(3)}$ are responsible for the $y$ component of the polarization, and in an isotropic medium, far away from resonances,
\begin{multline}
\frac{4\pi}{c^2} \frac{\partial^2(\bPel)_y}{\partial t^2} = \frac{2 n_0 n_2 \omega_p}{3c^2}  \left(   - \omega_p I_e A_y  \right. \\ \left. -2i \frac{\partial I_e}{\partial t} A_y-2i I_e \frac{\partial A_y}{\partial t} \right) e^{i\bkp \cdot \br -i\omega_p t}.
\label{finalKerr2}
\end{multline}
Details are given in Appendix A.

Putting Eq.~(\ref{finalKerr}) in Eq.~(\ref{wave3}), we have
\begin{equation*}
\frac{\partial A_x}{\partial z} + \frac{n_0}{c}\frac{\partial A_x}{\partial t} = \frac{2 n_2}{c} \left( i  \omega_p I_e - 2\frac{\partial I_e}{\partial t} \right) A_x - \frac{4 n_2}{c} I_e \frac{\partial A_x}{\partial t},
\end{equation*}
where we have used $\omega_p/k_p = c/n_0$.
Defining $A_x = |A_x| e^{i\Phi_x}$, one can show that the nonlinear change in phase as the probe propagates through the interaction region is
\begin{equation}
\frac{\partial \Phi_x}{\partial z} - \frac{n_0}{c}\frac{\partial \Phi_x}{\partial t}  = k_p (2n_2 I_e).
\label{phasex}
\end{equation}
Thus, the effective Kerr nonlinear refractive index for the probe is found to be $\Delta n^{\mathrm{cross}}_K = 2n_2 I_e$, twice as large as the change in nonlinear index calculated for the pump beam on itself, $\Delta n_K = n_2 I_e$.
The phase shift of the probe is enhanced by a factor of two regardless of whether the probe is degenerate or nondegenerate with the pump.
Similar calculations using higher-order response functions predict enhanced probe phase shifts for the higher-order Kerr effect due to $\chi^{(5)}$, $\chi^{(7)}$, etc.~\cite{loriot_measurement_2010}.

The factor of two enhancement of the Kerr phase shift of a weak probe due to cross phase modulation compared with the phase shift of the pump due to self phase modulation can be explained physically by diffraction from a nonlinear index grating \cite{smolorz_femtosecond_2000,wahlstrand_effect_2011}.
To understand this, it is helpful to consider the problem in a different, less rigorous way.
The total intensity is, in the limit of a weak probe,
\begin{equation}
I(\br,t) \cong \frac{n_0 c}{8\pi} \left[ |A_e|^2  + \left(A_e^* A_x e^{i \Delta \bk \cdot\br-i\Delta \omega t} + c.c. \right) \right],
\label{intensity}
\end{equation}
where $\Delta \bk = \bk_p - \bk_e$ and $\Delta \omega = \omega_p - \omega_e$.
Interference of the pump and probe beams causes a sinusoidal modulation of the total intensity (the terms in parenthesis) where the beams cross.
When inserted into $\Delta n_K = n_2 I$, we find $\Delta n_K = \Delta n^s_K + (\Delta n^g_K e^{i\Delta \bk \cdot \br - i\Delta \omega t}+c.c.)$, where $\Delta n^s_K = n_2 I_e$ is the ``smooth'' nonlinear index and $\Delta n^g_K = n_2 n_0 c/(8\pi) A_e^* A_x$ is the ``grating'' nonlinear index.
The $x$ component of the polarization due to this nonlinear refractive index is $(\bPel)_x = [\Delta n_K n_0/(2\pi)] E_x$.
The terms that contain $e^{i\bkp \cdot \br - i\omega_p t}$ are
\begin{eqnarray}
(\bPel)_x &=& \frac{n_0}{4\pi} \left( \Delta n^s_K A_x e^{i\bkp \cdot \br - i\omega_p t} \right. \nonumber \\ && \left. + \Delta n^g_K e^{i\Delta \bk \cdot \br - i\Delta \omega t} A_e e^{i\bke \cdot \br - i\omega_e t} \right) \label{Kerrgrating} \\
&=& \frac{n_0}{4\pi} \left( n_2 I_e A_x e^{i\bkp \cdot \br - i\omega_p t} \right. \nonumber \\ && \left. + n_2 \frac{n_0 c}{8\pi} |A_e|^2 A_x e^{i(\bke+\Delta \bk) \cdot \br - i(\omega_e+\Delta \omega) t} \right) \nonumber \\
&=& \frac{n_0}{4\pi} \left( n_2 I_e A_x e^{i\bkp \cdot \br - i\omega_p t} + n_2 I_e A_x e^{i\bkp \cdot \br - i\omega_p t} \right). \nonumber
\end{eqnarray}
Taking the second derivative with respect to time and applying the SVEA leads to Eq.~(\ref{finalKerr}).
The smooth component $\Delta n^s_K$ produces the same Kerr phase shift in the probe field as the pump induces on itself, and the grating component $\Delta n^g_K$ combines with the pump field to produce an identical phase shift in the probe field, doubling the overall response.

Pump light is diffracted by the nonlinear index grating into the probe path with just the right phase and amplitude to produce a doubling in the probe phase shift.
Probe light is also diffracted into the pump beam direction, and no net energy transfer occurs.
The extra probe phase shift from the grating is independent of the magnitude of the probe field, since the index modulation scales with the probe field.
When $\omega_e \neq \omega_p$, the Kerr grating moves in time as well as space.
The diffracted pump light at frequency $\omega_e$ is frequency shifted to exactly match the probe light at frequency $\omega_p$, and the factor of two enhancement in the phase shift still occurs.
Note that the crossing angle $\theta$ does not appear in Eqs.~(\ref{finalKerr},\ref{finalKerr2}); for $\omega_e \neq \omega_p$ the same factor of two enhancement in the probe phase shift occurs even for $\theta = 0$, where there is no transverse spatial grating.

\subsection{Plasma nonlinearity}
Ionization of atoms and molecules by the intense optical field frees electrons, which have a strong negative polarizability.
This causes the refractive index to change with time during the pump pulse.
We calculated cross phase modulation effects from this plasma response previously assuming a multiphoton ionization model \cite{wahlstrand_effect_2011}.
Here we generalize to the case of an ionization rate $w(I)$ that depends only on the cycle averaged intensity, $I$.
Simple ionization models exist that allow coverage of the multiphoton and tunneling limits \cite{perelomov__1966,popruzhenko_strong_2008}, as well as the intermediate intensity regime where laser filamentation typically occurs in gases.
Note that our treatment here neglects harmonic generation, including that arising from the time-dependent ionization rate within an optical cycle (for example, \cite{brunel_1990,yudin_2001,verhoef_2010}).
As mentioned previously, harmonic generation leads to an effective nonlinear refractive index through cascading \cite{bache}, but it is negligible for the experimental geometries considered.

The presence of free electrons leads to a polarization term
\begin{equation*}
\frac{\partial^2 \bP_{pl}}{\partial t^2} = \frac{e^2}{m_e} N_e (\br, t) \bE (\br,t),
\end{equation*}
where $N_e$ is the electron density.
The ionization rate is a function of time and space, and yields the electron density through $dN_e/dt = N_a w \left(I(\br,t) \right)$, where $N_a$ is the number of atoms or molecules per cm$^{3}$.
Using the total intensity given by Eq.~(\ref{intensity}) we have
\begin{multline}
w(I(\br,t)) \cong w(I_e(\br,t)) \\ + \left. \frac{n_0 c}{8\pi} \left[ \frac{dw}{dI}\right|_{I_e(\br,t)} A_e^*(\br,t) A_x(\br,t) e^{i\Delta\bk \cdot\br-i\Delta \omega t}+c.c.\right]. \nonumber
\end{multline}

In analogy with the discussion in the previous section, the total free electron density can be split into smooth and grating terms $N_e=N_e^s+(N_e^g e^{i\Delta \bk \cdot\br}+c.c.)$, where the smooth term is
\begin{equation*}
N_e^s (\br, t) = N_a \int_{-\infty}^t w(I_e(\br,t')) dt',
\end{equation*}
and the grating term is 
\begin{multline}
N_e^g (\br, t) = \\ \frac{n_0 c}{8\pi} N_a \int_{-\infty}^t \left. \frac{dw}{dI}\right|_{I_e(\br,t')} A_e^*(\br,t') A_x(\br,t') e^{-i\Delta \omega t'} dt'.
\label{Negrat}
\end{multline}
Note that we have assumed the weak ionization limit (\emph{i.e.}~neglecting depletion of the neutral atom population), which is a good approximation for the experiments we compare to.
The integral in Eq.~(\ref{Negrat}) is suppressed when $\omega_p \neq \omega_e$.

As described previously, we are interested only in the terms containing $e^{i\bk_p \cdot \br}$, as only they contribute to the signal in the probe beam direction.
These are
\begin{multline}
\frac{\partial^2 \bP_{pl}}{\partial t^2} = \frac{e^2}{m_e} \left[ N_{e}^s (\br, t) \bAp (\br,t) e^{i\bkp \cdot \br-i\omega_p t} \right. \\ \left. + N_{e}^g (\br, t) \bAe (\br,t)  e^{i\bkp \cdot \br-i\omega_e t} \right].
\label{source_plasma}
\end{multline}
The smooth term causes a phase shift in the probe pulse, consistent with a change in index $\Delta n = -N_e/2N_{\mathrm{cr}}$, where $N_{\mathrm{cr}}$ is the critical density.
The pump beam combines with the grating term $N_e^g$ to create a polarization source propagating in the probe direction.
In the calculations shown later we shall find that the main result of the grating term is a negative phase shift for the probe polarization component parallel to the pump polarization \cite{wahlstrand_effect_2011}.
The grating phase shift only appears when $\omega_p \approx \omega_e$.

\subsection{Rotational nonlinearity}
In molecular gases, an intense laser field applies a torque to the molecules, partially aligning them with the optical field \cite{lin_birefringence_1976,morgen,seideman,stapelfeldt,ripoche_determination_1997}.
The effective third-order nonlinearity due to this alignment can be an important, even dominant part of the total nonlinearity in air for intensities below the ionization threshold \cite{chen_single-shot_2007,wahlstrand_absolute_2012}.
The nonlinear rotational response comes about because in diatomic molecules the polarizability depends on the angle between the laser field direction and the molecular axis.
For an angle $\theta$ between the laser polarization and the molecular axis, the effective polarizability can be written as $\alpha_{\mathrm{eff}} = \alpha_{\perp} + \Delta \alpha \cos^2 \theta$, where $\alpha_\perp$ is the polarizability for the optical field perpendicular to the molecular axis and $\Delta \alpha = \alpha_\parallel - \alpha_\perp$ is the polarizability anisotropy, the difference in polarizability for light polarized parallel and perpendicular to the molecular axis.
In a linearly polarized optical field, the molecules tend to align into the field in order to minimize their energy, but because of rotational inertia, there is a time lag.
With a strong optical field, the alignment of molecules causes the linear susceptibility to change as a function of time, leading to an effective odd-order nonlinearity.
To lowest order the rotational response is linear in the pulse energy.
The case of a linearly polarized field has been considered in detail previously \cite{nibbering_determination_1997,chen_single-shot_2007}.

For pump and probe beams of differing polarization, it will be shown here that the calculation has additional complexity.
The molecular orientation-dependent dipole moment induced by the optical field is $\bf{p} = \alpha \bE$, where $\alpha$ is the polarizability tensor.
Since the pump and probe fields are assumed to be polarized in the $xy$ plane, we only need consider the components of the induced dipole moment in that plane.
These are
\begin{subequations}
\label{dipole2}
\begin{eqnarray}
p_x  &=& (\alpha_\perp + \Delta \alpha \cos^2 \theta) [\bEe(t) + \bEp(t)]_x \nonumber \\ && + \Delta \alpha \sin \theta \cos \theta \cos \phi [\bEp(t)]_y   \\
p_y  &=& (\alpha_\perp + \Delta \alpha \sin^2 \theta \cos^2\phi) [\bEp(t)]_y \nonumber \\ && + \Delta \alpha \sin \theta \cos \theta \cos \phi [\bEe(t) + \bEp(t)]_x,
\end{eqnarray}
\end{subequations}
where $\theta$ is the angle between the molecular axis and the $x$ direction, and $\phi$ is the azimuthal angle about the $x$ axis, measured with respect to the $y$ direction.
For brevity, we omit the $\br$ argument in this section -- note that all quantities with a time dependence also depend on $\br$.
The polarization of the gas is $\bP(t) = N_a \langle \mathbf{p} \rangle_t$, where $N_a$ is the number density of the gas and $\langle \rangle_t$ denotes the time-dependent ensemble average.
We find
\begin{eqnarray}
P_x  &=& N_a \left(\alpha_\perp + \Delta \alpha \langle \cos^2 \theta\rangle_t \right) [\bEe(t) + \bEp(t)]_x \nonumber \\ && + \frac{N_a}{2}  \Delta \alpha \langle \sin 2\theta \cos \phi \rangle_t [\bEp(t)]_y  \nonumber \\
P_y  &=& N_a  \left[\alpha_\perp + \frac{\Delta \alpha}{2} \left(1-\langle\cos^2 \theta \rangle_t \right) -\frac{\Delta\alpha}{2}  \langle \cos^2 \theta \cos 2\phi \rangle_t   \nonumber \right. \\ && \left. +\frac{\Delta\alpha}{2} \langle \cos 2\phi \rangle_t \right]  [\bEp(t)]_y \nonumber \\ &&+ \frac{N_a\Delta \alpha}{2} \langle \sin 2\theta \cos \phi \rangle_t [\bEe(t) + \bEp(t)]_x. \nonumber
\end{eqnarray}
For the calculation we are only interested in the nonlinear response, which excludes the static background refractive index proportional to $N_a(\alpha_\perp+ \Delta \alpha/3)$.
Removing this static contribution, we find that the nonlinear rotational polarization is
\begin{subequations}
\begin{eqnarray}
(\bProt)_x &=& N_a \Delta \alpha \left\{ \left( \langle\cos^2 \theta \rangle_t -\frac{1}{3} \right) [\bEe (t) + \bEp(t)]_x \right. \nonumber \\ && \left. +\frac{1}{2} \langle \sin 2\theta \cos \phi \rangle_t [\bEp(t)]_y \right\}, \\
(\bProt)_y &=& N_a \Delta \alpha \left\{ -\frac{1}{2} \left[ \left( \langle \cos^2 \theta \rangle_t -\frac{1}{3} \right) \right. \right. \nonumber \\ && \left. \left. + \langle \cos^2 \theta \cos 2\phi \rangle_t - \langle \cos 2\phi \rangle_t \right][\bEp(t)]_y \right. \nonumber \\ 
&& \left. +\frac{1}{2}  \langle \sin 2\theta \cos \phi \rangle_t [\bEe (t) + \bEp(t)]_x \right\}.
\end{eqnarray}
\end{subequations}

The Hamiltonian for the interaction of an optical field $\bE$ and the rotational modes of molecules is
\begin{math}
H = -(1/2)\bf{p} \cdot \bE.
\end{math}
Using Eqs.~(\ref{dipole2}), we find, to lowest order in the probe field and keeping only terms that vary slowly compared with $\omega_e$ and $\omega_p$, $H=H^{xx}_\perp+H^{xx}_\parallel+H^{xy}$, where
\begin{subequations}
\begin{eqnarray}
H^{xx}_\perp (t) &=& -\frac{1}{2} \alpha_\perp \left( |A_e|^2 \right. \nonumber \\ && \left. + [ A_e^* A_x e^{i\Delta \bk \cdot \br - i\Delta \omega t} +c.c.] \right), \\
H^{xx}_\parallel (t) &=& -\frac{1}{2} \Delta \alpha \cos^2 \theta \left( |A_e|^2 \right. \nonumber \\ && \left. + [ A_e^* A_x e^{i\Delta \bk \cdot \br - i\Delta \omega t} +c.c.]\right), \label{Hxx} \\
H^{xy} (t) &=& -\frac{1}{4} \Delta \alpha \sin 2\theta \cos \phi \nonumber \\ && \times [ A_e^* A_y e^{i\Delta \bk \cdot \br - i\Delta \omega t}+c.c.]. \label{Hxy}
\end{eqnarray}
\end{subequations}
To lowest order, the rotational contribution to the nonlinear polarization is third-order in the optical field, and we restrict ourselves here to this case.
We seek a perturbative solution for the time evolution of the density matrix $\bm{\rho}$ using the basis of rotational eigenstates $|j,m\rangle$, having quantum numbers $j$ for the total rotational angular momentum $\bf{J}$ and $m$ for the component of $\bf{J}$ along the $x$ direction.
The zeroth order solution $\bm{\rho}^{(0)}$ describes the initial thermal equilibrium distribution of rotational states.
It is diagonal and depends only on the total angular momentum $j$,
\begin{equation}
\rho^{(0)}_{(jm),(j'm')} = \frac{D_j  e^{-\frac{E_j}{k_B T}}}{\sum_k D_k (2k+1) e^{-\frac{E_k}{k_B T}}} \delta_{jj'} \delta_{mm'},
\label{rho0}
\end{equation}
where $E_j=hcBj(j+1)$ is the energy of the $j^{\mathrm{th}}$ rotational level ($B$ is a rotational constant, which is related to the molecular moment of inertia), and $D_j$ is a degeneracy factor related to nuclear spin statistics \cite{nibbering_determination_1997,chen_single-shot_2007}.
For brevity we define $\rho^{(0)}_{jm} = \rho^{(0)}_{(jm),(jm)}$.
The first order solution is
\begin{multline}
\rho^{(1)}_{(jm)(j'm')} (t) = \\ -\frac{i}{\hbar} \int_{-\infty}^t [\mathbf{H}(t'), \bm{\rho}^{(0)}]_{(jm),(j'm')} e^{i\omega_{jj'} (t'-t)} dt',
\label{rho1}
\end{multline}
where [,] denotes the commutator and $\omega_{jj'} = (E_j-E_{j'})/\hbar$.
We neglect decay of the rotational coherences, as we are interested here in the response during the pump pulse and a few hundred femtoseconds afterward, whereas the rotational coherence lasts $>10$ ps at atmospheric pressure and temperature \cite{chen_single-shot_2007}.

Since $\bm{\rho}^{(0)}$ and $\mathbf{H}^{xx}_\perp$ are diagonal, $[\mathbf{H}, \bm{\rho}^{(0)}]_{(jm),(j'm')}  = (\rho^{(0)}_{jm}-\rho^{(0)}_{j'm'}) [H^{xx}_\parallel]_{(jm),(j'm')}+(\rho^{(0)}_{jm}-\rho^{(0)}_{j'm'}) H^{xy}_{(jm),(j'm')}$, where
\begin{eqnarray}
H^{xx}_{\parallel,(jm),(j'm')}(t) &=& -\frac{1}{2} \Delta \alpha \left[ |A_e|^2 \nonumber \right. \\ && \left. + \left(A_e^* A_xe^{i\Delta \bk \cdot \br - i\Delta \omega t} + c.c. \right) \right] \nonumber \\ && \times \langle jm | \cos^2 \theta | j'm' \rangle \nonumber \\
H^{xy}_{(jm),(j'm')}(t) &=& -\frac{1}{4} \Delta \alpha \left(A_e^* A_y e^{i\Delta \bk \cdot \br - i\Delta \omega t} +c.c. \right) \nonumber \\ && \times \langle jm | \sin 2\theta \cos \phi | j'm' \rangle. \nonumber
\end{eqnarray}
In Appendix B expressions for $\langle jm | \cos^2 \theta | j'm' \rangle$ and $\langle jm | \sin 2\theta \cos \phi | j'm' \rangle$ are given.
Finally, to find the nonlinear polarization, we need to calculate
\begin{multline*}
\langle \sin 2\theta \cos \phi \rangle_t = \mathrm{Tr} \left[\bm{\rho}^{(1)} (t) \sin 2\theta \cos \phi \right] = \\ \sum_{j,m,j',m'} \rho^{(1)}_{(jm),(j'm')} (t) \langle j,m| \sin 2\theta \cos \phi | j',m' \rangle,
\end{multline*}
and the equivalent expression for $\langle \cos^2\theta \rangle_t$ \cite{chen_single-shot_2007}.
One can show that to lowest order $\langle \cos^2 \theta \cos 2\phi \rangle_t = \langle \cos 2\phi \rangle_t = 0$.
In Appendix B, it is shown that, in analogy with our previous discussion, the ensemble-averaged molecular alignment quantities can be split into smooth and grating terms according to $\langle \cos^2 \theta \rangle_t = \langle \cos^2 \theta \rangle^s_t +1/3 + (\langle \cos^2 \theta \rangle^g_t e^{i\Delta \bk \cdot \br} + c.c.)$ and $\langle \sin 2\theta \cos \phi \rangle_t = \langle \sin 2\theta \cos \phi \rangle^g_t e^{i\Delta \bk \cdot \br} + c.c.$, where
\begin{eqnarray}
\langle \cos^2 \theta \rangle^s_t &=& \sum_j K_j \int_{-\infty}^t \sin[\omega_{j+2,j} (t'-t) ] \nonumber \\ &&\times |A_e|^2 (t') dt', \label{cos2thetast} \\
\langle \cos^2 \theta \rangle^g_t &=& \sum_j K_j \int_{-\infty}^t \sin[\omega_{j+2,j} (t'-t) ] \nonumber \\ &&\times A_e^* (t') A_x (t') e^{-i \Delta \omega t'} dt', \label{cos2thetagt} \\
\langle \sin 2\theta \cos \phi \rangle^g_t &=& \frac{3}{2} \sum_j K_j \int_{-\infty}^t \sin[\omega_{j+2,j} (t'-t) ] \nonumber \\ &&\times A_e^* (t') A_y ( t') e^{-i \Delta \omega t'} dt', \label{sin2thetacosphit}
\end{eqnarray}
and
\begin{equation*}
K_j = \frac{2\Delta \alpha}{15\hbar} \frac{(j+1)(j+2)}{2j+3} \left( \frac{\rho^{(0)}_{j+2}}{2j+5} -\frac{\rho^{(0)}_{j}}{2j+1}\right),
\end{equation*}
where $\rho^{(0)}_{j} \equiv (2j+1) \rho^{(0)}_{jm}$.
As with the plasma response, the integrals in the grating terms in Eq.~(\ref{cos2thetagt}) and Eq.~(\ref{sin2thetacosphit}) are suppressed when $\Delta \omega$ is nonzero.
However, resonant behavior may occur when the pump and probe frequencies differ by the spacing between rotational levels.
In the experiments considered where $\Delta \omega \neq 0$, $\Delta \omega$ is very large compared to the spacing between thermally populated rotational levels.

Keeping only terms that lead to a polarization source propagating in the $e^{i\bkp \cdot \br}$ direction, we find
\begin{eqnarray}
(\bProt)_x &=& N_a \Delta \alpha  \left( \langle \cos^2 \theta \rangle^s_t A_x(t) e^{i\bkp \cdot \br - i\omega_p t} \nonumber \right. \\ && \left. + \langle \cos^2 \theta \rangle^g_t A_e (t) e^{i\bkp \cdot \br - i\omega_e t} \right) , \nonumber \\
(\bProt)_y &=& N_a \Delta \alpha \left( -\frac{1}{2}\langle \cos^2 \theta \rangle^s_t  A_y (t) e^{i\bkp \cdot \br - i\omega_p t} \nonumber \right. \\ && \left. +\frac{1}{2} \langle \sin 2\theta \cos \phi \rangle^g_t A_e (t) e^{i\bkp \cdot \br - i\omega_e t} \right). \nonumber
\end{eqnarray}
Taking the second derivative with respect to $t$, we find (in the SVEA and keeping only terms without time derivatives)
\begin{subequations}
\begin{eqnarray}
\frac{\partial^2 (\bP_{\mathrm{rot}})_x}{\partial t^2} &=& -N_a \Delta \alpha \omega_p^2 \left[ \langle \cos^2 \theta \rangle^s_t  A_x (t) e^{- i\omega_p t} \right. \nonumber \\&& \left. +\frac{\omega_e^2}{\omega_p^2} \langle \cos^2 \theta \rangle^g_t A_e (t) e^{- i\omega_e t} \right], \\
\frac{\partial^2 (\bP_{\mathrm{rot}})_y}{\partial t^2} &=& \frac{1}{2}N_a \Delta \alpha \omega_p^2 \left[ \langle \cos^2 \theta \rangle^s_t  A_y (t) e^{- i\omega_p t} \right. \nonumber \\ && \left.  - \frac{\omega_e^2}{\omega_p^2} \langle \sin 2\theta \cos \phi \rangle^g_t A_e (t) e^{- i\omega_e t} \right].
\end{eqnarray}
\label{source_rotational}
\end{subequations}
It will be shown later that the rotational grating produces a probe phase shift that approximately follows the pump envelope.
In the first-order perturbation approximation used here, the rotational grating contribution is proportional to the pump intensity \cite{secondrot}.

%In the limit of a long optical pulse, the rotational nonlinearity contributes to an effective $n_2$. Long pulse limit \cite{erratum}.

\section{Calculations}
We next perform numerical calculations and compare them to previously reported experimental results.
Defining a local time coordinate $\tau = t-zn_0/c$, so that $\partial/\partial z \rightarrow \partial /\partial z-(n_0/c)\partial/\partial\tau$ and $\partial/\partial t \rightarrow \partial/\partial \tau$ \cite{boyd}, and defining barred functions using this local coordinate system $\bbAp(\br,\tau) = \bAp(\br, t)$, Eq.~(\ref{wave3}) becomes
\begin{equation}
i k_p \frac{\partial\bbAp}{\partial z} e^{-i\omega_p \tau} = \frac{4\pi}{c^2} \frac{\partial^2\bar{\bP}_{\mathrm{NL},p}}{\partial \tau^2},
\label{wave4}
\end{equation}
Note that when these coordinates are used, the probe envelope is independent of $z$ in the absence of nonlinear interaction (i.e. when the right hand side is zero).
The outgoing probe field is first calculated numerically using Eq.~(\ref{wave4}), with the nonlinear polarization sources from Eqs.~(\ref{finalKerr},\ref{finalKerr2},\ref{source_plasma},\ref{source_rotational}).
The signal measured in two types of experiments is then calculated from the probe field.

For the ionization rate $w(I)$ we use the model of Popruzhenko \emph{et al.}~\cite{popruzhenko_strong_2008}, which has been shown to approximately agree with time domain Schr\"odinger Equation calculations and has been used in recent Kramers-Kronig calculations of the nonlinear response \cite{wahlstrand_high_2012,bree_saturation_2011}.
For the rotational response in N$_2$ and O$_2$, we use $B=2.0$ cm$^{-1}$ for N$_2$ and $B=1.44$ cm$^{-1}$ for O$_2$ and assume $T=293$ K.
For $n_2$ and $\Delta \alpha$ we use the values reported in \cite{wahlstrand_absolute_2012}.
We neglect the vibrational response, which in N$_2$ and O$_2$ results in a small (less than 10\% of the electronic nonlinearity \cite{shelton_measurements_1994}) effectively instantaneous response for the pulse durations used in the experiments.

\subsection{Single-shot spectral interferometry: Degenerate vs. nondegenerate probe}
We first simulate the signal in spectral interferometry experiments, which directly measure the time- and space-dependent probe phase shift.
We calculate using the parameters of recent experiments using a chirped probe pulse \cite{wahlstrand_absolute_2012,wahlstrand_high_2012}.
The pump pulse is assumed to be transform limited and Gaussian, centered at 800 nm with a full width at half maximum (FWHM) time duration of 40 fs.
The linearly polarized probe is either centered at 800 nm (degenerate) or 600 nm (nondegenerate).
The chirped probe pulse is handled with a time-dependent phase in the incident complex probe envelope $A_p^{\mathrm{inc}}(\br, t)$.
For the nondegenerate case, the outgoing probe field calculated is frequency filtered (as in the experiment) to remove components near the pump frequency $\omega_e$ before the nonlinear phase shift $\Delta\Phi(t) = \mathrm{Im}\{\ln[A_{p}(t)/A_{p}^{\mathrm{inc}}(t)]\}$ is calculated.

\begin{figure}
\begin{center}
\includegraphics[width=8.5cm]{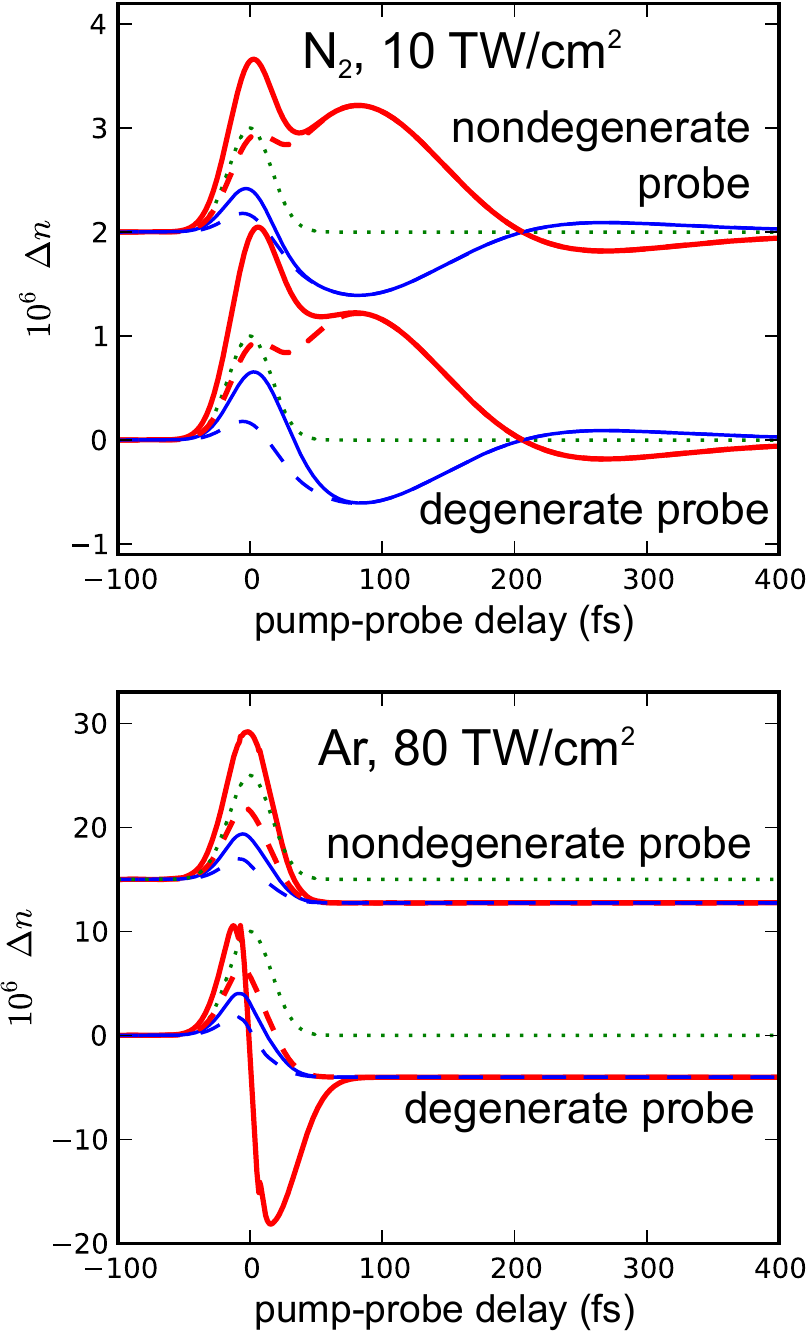}
\end{center}
\caption{(color online) Simulated single-shot spectral interferometry signal in N$_2$ and Ar for a transform-limited, 40 fs pump pulse centered at 800 nm. The pump intensity is shown as a green dotted line.
The calculated nonlinear index $\Delta n (t) = \Delta \Phi(t) c/(n_0 \omega_p L)$, where $\Phi(t)$ is the time-dependent probe phase shift, is shown assuming a nondegenerate probe centered at 600 nm (upper curves) and a degenerate probe centered at 800 nm (lower curves). The change in index is plotted for the probe polarized parallel (thick red) or perpendicular (thin blue) to the pump polarization.
Curves for a nondegenerate probe are offset vertically for clarity.
The simulated signal is shown including (solid) and excluding (dashed) grating effects.
}
\label{sssi}
\end{figure}

The time-dependent nonlinear refractive index $\Delta n (t) = \Delta \Phi(t)c/(n_0 \omega_p L)$, where $L=200$ $\mu$m is the interaction length, is shown in Fig.~\ref{sssi}.
Two different calculations are shown: one for N$_2$ at low pump intensity, which shows the effect of the rotational response, and Ar at high pump intensity, which shows the effect of the plasma response.
The green dotted line shows the intensity envelope of the pump pulse, and the thick red (thin blue) curves show the response for the probe polarization parallel (perpendicular) to the pump polarization.
Solid and dashed lines show the simulated signal with and without the grating response, respectively.
The dashed line thus indicates the nonlinearity that a laser pulse would impart on itself.

In N$_2$ at intensities below the ionization threshold, the bound electronic nonlinearity due to $\chi^{(3)}$ causes a phase shift that is proportional to the pump intensity envelope, and the smooth rotational nonlinearity causes a delayed response peaking about 80 fs after the peak of the pump pulse.
When a degenerate probe is used (lower curves), an effectively instantaneous rotational grating signal appears that, if not properly taken into account in the analysis, would make it appear that $n_2$ was larger than it is.
Whether the probe polarization is oriented parallel or perpendicular to the pump polarization, the rotational grating contribution is positive, but not the same magnitude -- thus, it contributes to a transient birefringence experiment, as will be shown later.

In Ar at high intensity, the bound electronic nonlinearity is accompanied by the nonlinearity due to the generation of free electrons (plasma). 
The ``smooth'' plasma contribution accumulates during the pump pulse and afterward contributes a long-lived, constant phase shift.
The plasma grating results, for a degenerate probe polarized parallel to the pump polarization, in a strongly time dependent negative signal that is a few times larger than the ``smooth'' plasma signal.

In both cases, when a nondegenerate probe pulse is used (upper curves), the only difference between the experimentally measured $\Delta n$ and the true nonlinear index $\Delta n$ is the factor of two enhancement in the bound electronic contribution to the nonlinearity.
These calculations provide theoretical justification for the analysis performed in recent absolute measurements of the nonlinearity in the noble gases, N$_2$, O$_2$, and N$_2$O \cite{wahlstrand_absolute_2012,wahlstrand_high_2012}.
The extracted values for $n_2$ and $\Delta \alpha$ are in good agreement with previous calculations \cite{lin_birefringence_1976,maroulis_accurate_2003,bree_method_2010}, experiments using less direct methods \cite{shelton_measurements_1994,bridge_polarization_1966}, and a ratiometric measurement \cite{hertz_femtosecond_2000}.
Our calculations reinforce the point that nondegenerate supercontinuum spectral interferometry is the most direct, powerful technique currently available for measuring the optical nonlinearity.

\subsection{Transient birefringence using a degenerate probe}
\begin{figure}
\begin{center}
\includegraphics[width=8.5cm]{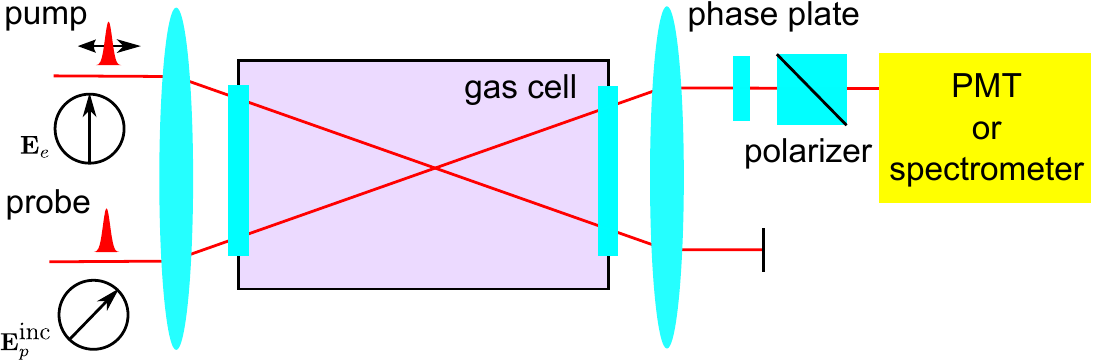}
\end{center}
\caption{(color online) Experimental diagram for transient birefringence experiment, showing the initial polarization of the pump and probe fields, the interaction region inside a gas cell, and the detection scheme.}
\label{loriot}
\end{figure}

Next we simulate the  degenerate pump-probe transient birefringence experiment by Loriot \emph{et al.} \cite{loriot_measurement_2009,loriot_negative_2011}.
A diagram of the heterodyne transient birefringence experimental apparatus is shown in Fig.~\ref{loriot}.
The probe is linearly polarized at 45$^\circ$ with respect to the pump polarization, so $A^{\mathrm{inc}}_x = A^{\mathrm{inc}}_y$.
After the interaction region, the probe beam is passed through a phase plate, which provides a static retardance $2\xi$ between the $x$ and $y$ probe components, followed by a polarizer oriented at $90^\circ$ with respect to the initial probe beam polarization.
The optical field after the polarizer is
\begin{multline}
A_{s} (\xi,t) = e^{i\xi} A_x (t) - e^{-i\xi} A_y (t) = \\ e^{i\xi} |A_x (t)|e^{i\Phi_x(t)} - e^{-i\xi} |A_y (t)|e^{i\Phi_y(t)}.
\label{heterodyne}
\end{multline}

In the Loriot \etal experiment, a photomultiplier tube is used to detect this field, and the signal is proportional to $\int |A_s (\xi,t)|^2 dt$.
The ``pure heterodyne'' signal is found by subtracting the signals for orientations of the phase plate producing static retardances $2\xi$ and $-2\xi$ \cite{loriot_measurement_2009},
\begin{eqnarray}
S_h &\propto& \int_{-\infty}^{\infty} |A_s(\xi,t)|^2 dt - \int_{-\infty}^{\infty} |A_s(-\xi,t)|^2 dt \nonumber \\ 
        &\propto& \sin(2\xi) \int_{-\infty}^{\infty} |A_x(t)||A_y(t)| \nonumber \\ &&\times \sin \left[\Phi_x(t)-\Phi_y(t) \right] dt,
\label{loriot_signal}
\end{eqnarray}
where $A_i(t) = |A_i(t)|e^{i\Phi_i(t)}$.
Since there is little pump-induced change in the probe amplitude, $|A_x| \approx |A_y|$, and $S_h$ is thus proportional to the pump-induced birefringence integrated with the probe intensity.
The heterodyne signal was measured as a function of the time delay between the pump and probe pulses.
We performed 3D numerical calculations of the Loriot experiment, assuming that the pump and probe beams were Gaussian beams with FWHM waists of 33 $\mu$m crossing at $\theta  = 4^\circ$.

\subsubsection{Rotational grating effects: simulations for N$_2$}
\begin{figure}
\begin{center}
\includegraphics[width=8.5cm]{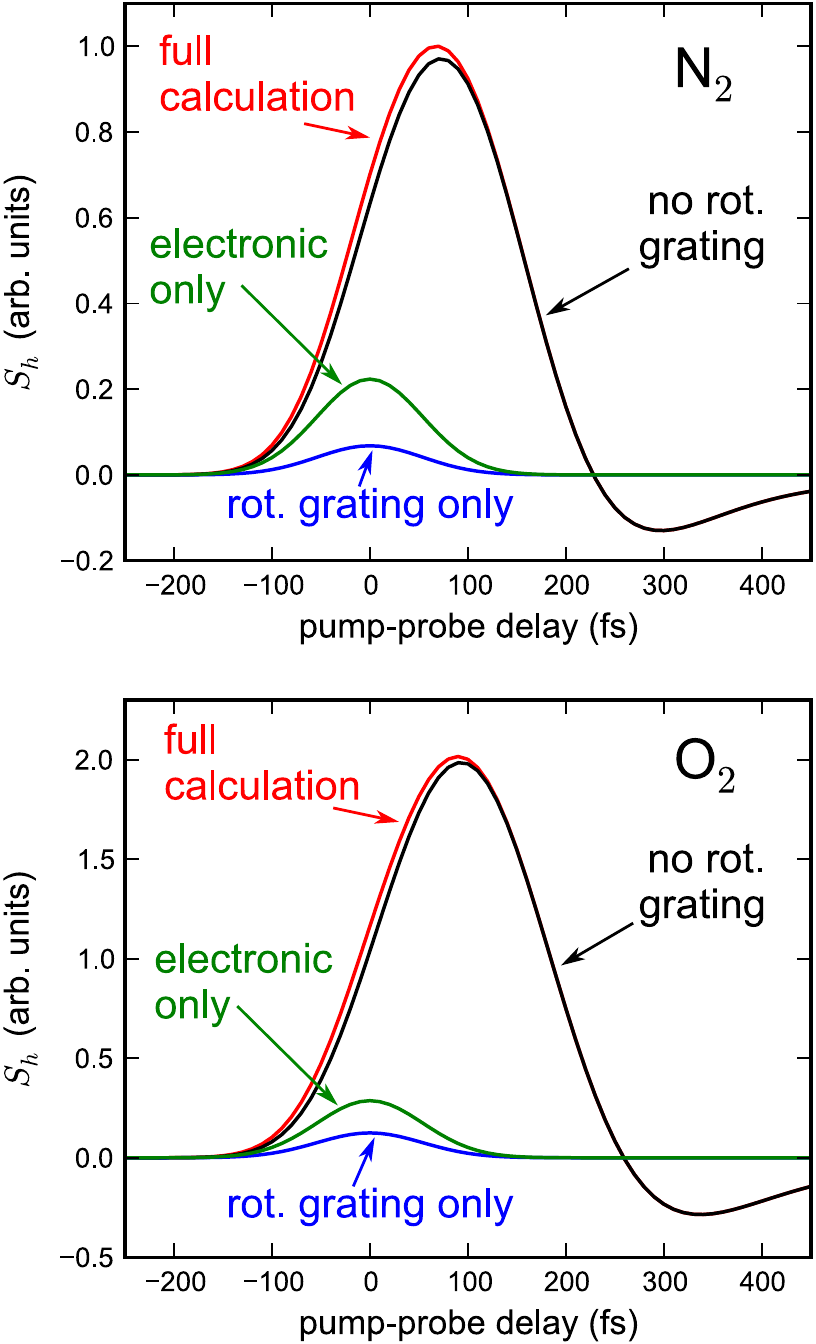}
\end{center}
\caption{(color online) Simulated heterodyne transient birefringence signal $S_h$ [Eq.~(\ref{loriot_signal})] for N$_2$ and O$_2$ as a function of pump-probe delay, assuming transform limited 90 fs pulses centered at 800 nm. 
The peak pump intensity is 10 TW/cm$^2$.
The curves show the calculated signal with and without the rotational grating contribution, the contribution from the bound electronic Kerr effect alone, and the contribution from the rotational grating alone.}
\label{loriot3}
\end{figure}

The calculated transient birefringence signal in N$_2$ and O$_2$ at low pump intensity is shown in Fig.~\ref{loriot3}, with and without the rotational grating contribution.
The rotational grating contribution alone is also shown, as is the electronic Kerr contribution alone. 
Both are roughly proportional to the convolved pump intensity envelope.
The rotational grating contribution in transient birefringence is positive, increasing the apparent instantaneous nonlinearity observed in a degenerate pump-probe experiment.

Table I shows measured values of $n_2$ from experiments by Loriot \emph{et al.}~\cite{loriot_measurement_2009} and from a recent nondegenerate spectral interferometry experiment by Wahlstrand \emph{et al.}~\cite{wahlstrand_absolute_2012}.
The reported values for Ar agree within error, but the values in N$_2$ and O$_2$ do not.
This can be explained by the additional phase shift imparted to the probe by the rotational grating in the Loriot \emph{et al.}~experiment (shown in Fig.~\ref{loriot3}).
We used the calculation to subtract the rotational grating contribution from the Loriot \emph{et al.}~measurements in N$_2$ and O$_2$; these corrected $n_2$ values are shown in the last column of Table 1.
After the correction the measurements agree within their uncertainty estimates.

\begin{table}
\begin{center}
\begin{tabular}{l | c | c | c }
Gas & \multicolumn{3}{c}{$n_2$ ($10^{-20}$ cm$^2$/W) }\\
& Ref.~\cite{wahlstrand_absolute_2012} & Ref.~\cite{loriot_measurement_2010} &  Ref.~\cite{loriot_measurement_2010} corrected for rot.~grating \\
\hline
Ar & $9.7\pm 1.2$ & $10.0\pm 0.9$ & -- \\
N$_2$ & $7.4 \pm 0.9$ &$11.0 \pm 2.0$ & $8.4\pm 2.0$ \\
O$_2$ & $9.5\pm 1.2$ &$16.0\pm 3.5$& $11.1\pm3.5$ \\
\end{tabular}
\caption{Comparison of $n_2$ measurements by Wahlstrand \emph{et al.}~\cite{wahlstrand_absolute_2012} and Loriot \emph{et al.}~\cite{loriot_measurement_2009,loriot_measurement_2010}.
For N$_2$ and O$_2$, values of $n_2$ from the Loriot \emph{et al.}~experiment are shown with and without correcting for the effectively instantaneous contribution from the rotational grating, which was calculated assuming 90 fs, transform-limited Gaussian pulses.}
\end{center}
\end{table}

\subsubsection{Plasma grating effects: simulations for Ar}
\begin{figure}
\begin{center}
\includegraphics[width=8.5cm]{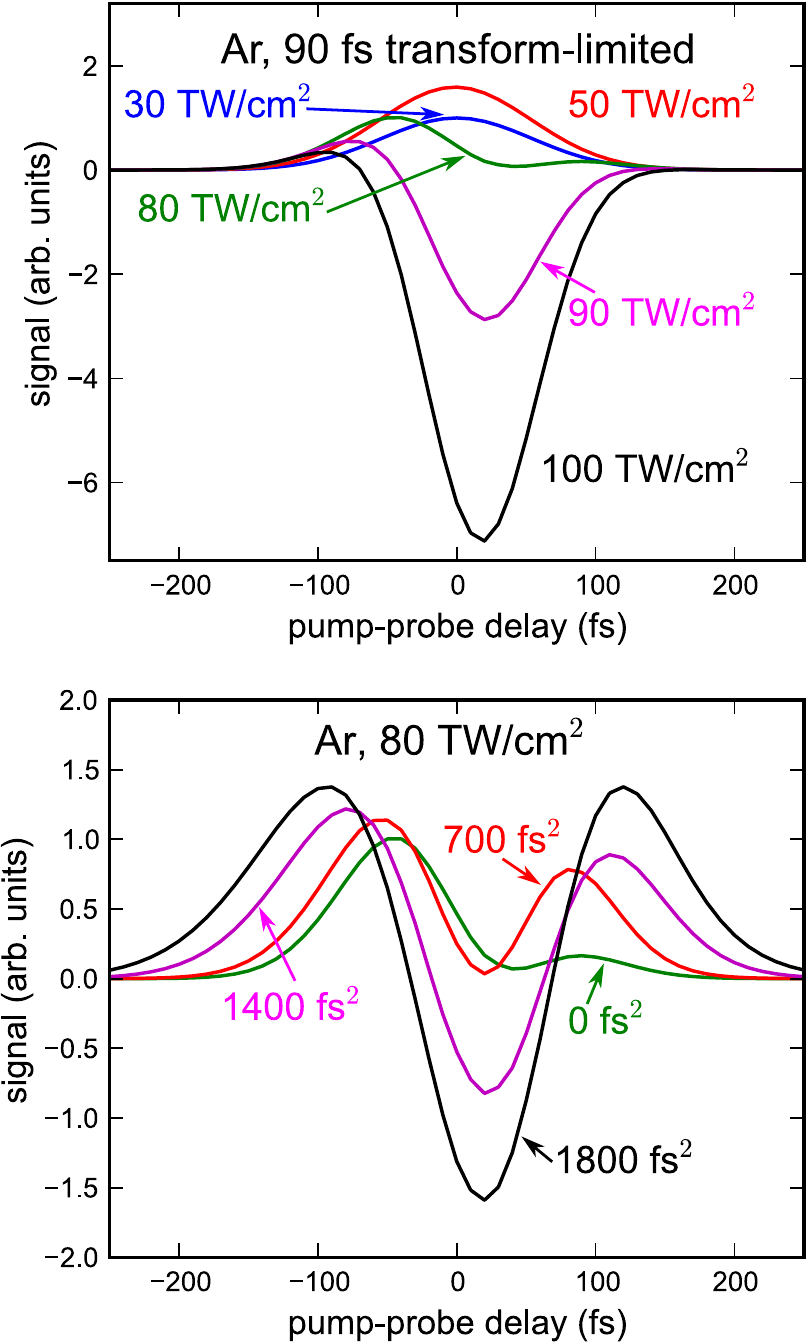}
\end{center}
\caption{(color online) Simulated heterodyne transient birefringence signal $S_h$ [Eq.~(\ref{loriot_signal})] for Ar as a function of pump-probe delay, assuming pulses centered at 800 nm.
In the upper panel, the calculation assumes a Gaussian transform limited 90 fs FWHM pulse, and the signal is shown for a few pump peak intensities.
In the lower panel, the calculation assumes a pump peak intensity of 80 TW/cm$^2$, and the signal is shown for a few values of group delay dispersion.
Both pump and probe pulses are chirped, and the transform-limited pulse width is 90 fs FWHM and Gaussian.
}
\label{loriot2}
\end{figure}

Plots of the calculated heterodyne transient birefringence signal versus pump-probe delay in Ar are shown in Fig.~\ref{loriot2}.
The signal for 90 fs transform-limited pump and probe pulses centered at 800 nm is shown as a function of pump intensity in the top panel of Fig.~\ref{loriot2}.
At low pump intensity the induced birefringence is $\Delta n_\parallel - \Delta n_\perp = 2n_2 - 2n_2/3 = 4n_2/3$, and the signal is proportional to the pump intensity envelope convolved with the probe intensity envelope.
At high pump intensity, this positive signal due to the Kerr birefringence is overwhelmed by the negative plasma grating signal.
The results are similar to those found previously from a one-dimensional calculation using a multiphoton ionization rate \cite{wahlstrand_effect_2011}.
We find that the negative birefringence due to the plasma grating appears at a similar intensity to that reported by Loriot \emph{et al.}~\cite{loriot_measurement_2009,loriot_negative_2011}.
Note that the normalized intensity used by Loriot \emph{et al.}~is the peak intensity divided by 1.7 \cite{loriot_measurement_2009,loriot_negative_2011}.

The plasma grating contribution dominates the signal at a lower peak intensity than one might expect from recent measurements using spectral interferometry, where the plasma phase shift first appears at approximately 80 TW/cm$^2$ \cite{wahlstrand_high_2012}.
The combination of three effects explain this apparent discrepancy.
First, the nonlinear refractive index change for a given plasma density is larger by a factor of $\sim 1.8$ for an 800 nm probe compared to a 600 nm probe owing to the $\lambda^2$ scaling of the free electron refractive index.
Thus, while a given plasma density produces a significantly larger phase shift at 800 nm compared to 600 nm,  the Kerr effect is nearly the same because of its comparatively lesser dispersion. 
Second, the calculation assumes 90 fs pulses, about 2.5 times longer than the pulses used in \cite{wahlstrand_optical_2011,wahlstrand_high_2012}, resulting in a higher plasma density for a given peak intensity.
Finally, the nearly instantaneous negative plasma grating signal is a few times larger than the ``smooth'' plasma phase shift that exists after the pump pulse (cf. Fig.~\ref{sssi}).
%The behavior reported by Loriot \emph{et al.}~seems to occur at an even lower intensity \cite{loriot_negative_2011}, but note that they plot the data as a function of a normalized pump intensity $\bar{I} = I/1.7$ that accounts for volume averaging effects.

The Loriot \emph{et al.} data appears roughly symmetric with respect to time reversal \cite{loriot_measurement_2009,loriot_negative_2011}.
Because the plasma grating signal is slightly delayed with respect to the Kerr signal due to the cumulative nature of the plasma nonlinearity, the signal for a transform limited pulse is asymmetric at intermediate intensity.
However, like two-beam coupling energy transfer, the plasma grating phase shift is sensitive to the chirp of the pump and probe pulses \cite{dogariu_purely_1997}.
If the pulses are chirped, the asymmetry of the pump-probe signal is lessened and can even be reversed.
This is illustrated in the bottom panel of Fig.~\ref{loriot2}, where the calculated signal is shown as a function of group delay dispersion (GDD), assuming that the transform-limited pulse width is 90 fs.
The pump and probe pulses are assumed to have the same GDD.

\section{Experiment}
Based on a numerical calculation in atomic hydrogen, B\'ejot \etal recently argued that the higher-order Kerr effect is present only at optical frequencies near the frequency of an intense driving field \cite{bejot_arxiv_2012}, and therefore nondegenerate pump-probe experiments, which use a probe pulse detuned from the pump pulse frequency, are incapable of observing it.
In this section we describe new experimental evidence that the negative, nearly instantaneous signal observed using a degenerate probe is caused by the plasma grating.

\subsection{Spectrally-resolved birefringence}
The Loriot \etal experiment measured the temporally-convolved transient birefringence as a function of pump-probe time delay \cite{loriot_measurement_2009}.
Another approach involves spectrally resolving the birefringence.
Previously, an experiment using 400 nm light in air was reported \cite{odhner_ionization-grating-induced_2012}, which showed that the intensity dependence of the negative signal was consistent with the plasma grating theory assuming multiphoton ionization \cite{wahlstrand_effect_2011}, which is a good approximation at 400 nm, where 4 photons are required for ionization of O$_2$.
Here, we report new data in Ar and air using 800 nm pulses.

\begin{figure*}
\begin{center}
\includegraphics[width=15cm]{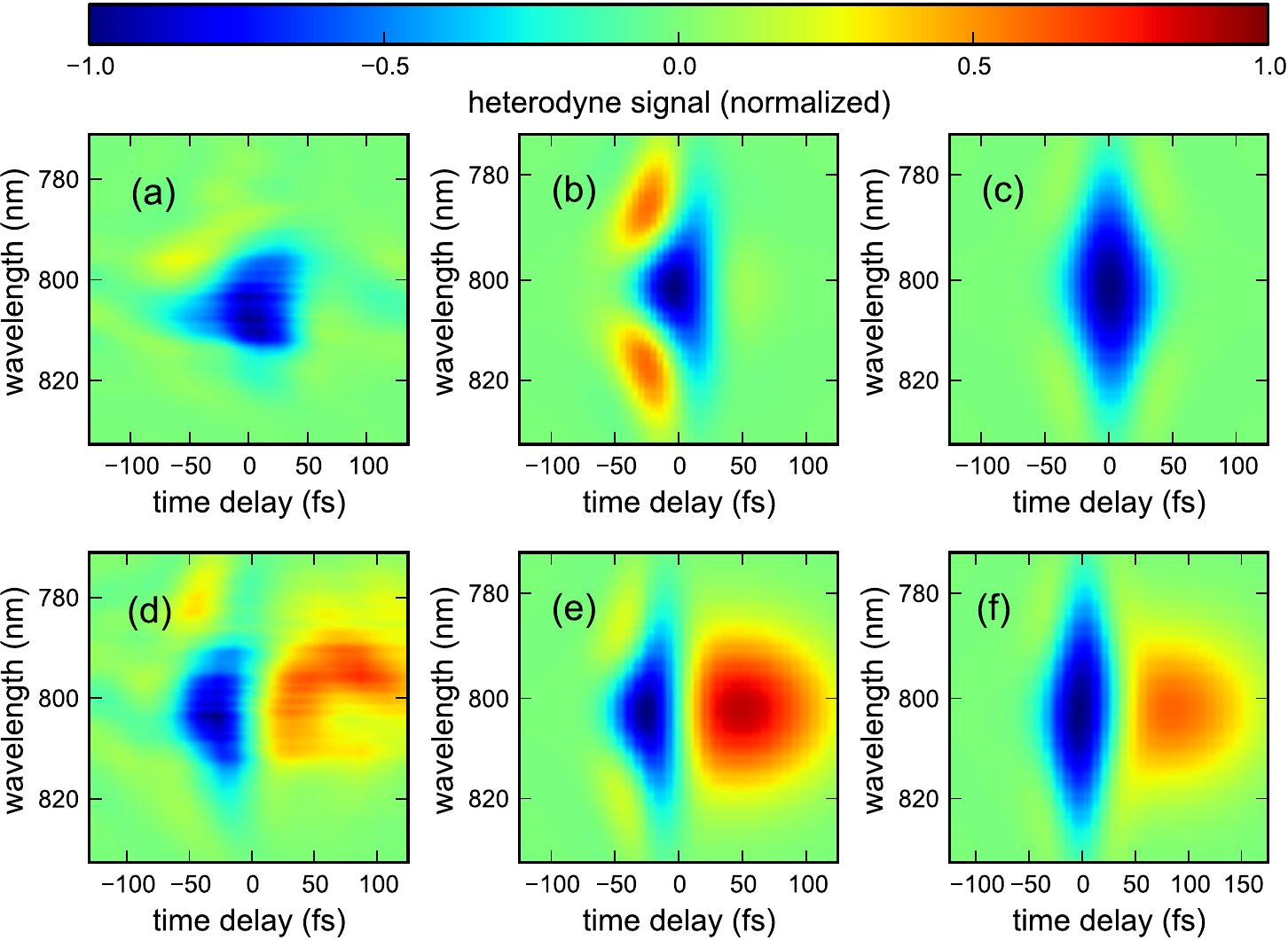}
\end{center}
\caption{(color online) Spectrally-resolved transient birefringence results.
The heterodyne-detected signal is plotted as a function of pump-probe time-delay and wavelength.
(a) Experimental data in Ar for pump pulse energy 42.5 $\mu$J.
(b) Calculation for Ar including plasma grating assuming peak intensity 100 TW/cm$^2$.
(c) Calculation for Ar assuming no plasma grating but a higher-order Kerr effect \cite{loriot_measurement_2010} using peak intensity 50 TW/cm$^2$.
(d) Experimental data in air for pump pulse energy 25 $\mu$J.
(e) Calculation for air including plasma grating assuming peak intensity 100 TW/cm$^2$.
(f) Calculation for air assuming no plasma grating but a higher-order Kerr effect \cite{loriot_measurement_2010} using peak intensity 60 TW/cm$^2$.
}
\label{trbir}
\end{figure*}

The experimental setup is similar to that shown in Fig.~\ref{loriot}.
After the pump and probe interact, a fixed quarter wave plate in the probe path produces a static birefringence $2\xi=\pi/4$, and then the probe beam passes through a polarizer.
The relative polarization of the pump beam with respect to the probe is alternated between $\pm 45^\circ$ to make measurements with $2\xi$ effectively $\pi/4$ and $-\pi/4$.
The spectrum of the beam after the polarizer is measured as a function of pump-probe time delay.
Measurements using $2\xi=\pi/4$ and $2\xi=-\pi/4$ are subtracted to generate the pure heterodyne spectrally-resolved pump-probe signal, with optical frequency on one axis and pump-probe time delay on the other.
The pump and probe were 42 fs FWHM, approximately transform limited pulses.
Experimental results are shown in Fig.~\ref{trbir}a for Ar and Fig.~\ref{trbir}d for air.

To simulate the experimental data, we numerically calculate the optical field at the output of the polarizer, given by Eq.~(\ref{heterodyne}), and then Fourier transform it and calculate the magnitude squared.
The pure heterodyne spectrum is found by subtracting the calculated Fourier transforms for $2\xi=\pi/4$ and $2\xi=-\pi/4$.
The signal is plotted as a function of optical frequency and pump-probe time delay in Fig.~\ref{trbir}b (Ar) and Fig.~\ref{trbir}e (air) for the plasma grating model and in Fig.~\ref{trbir}c (Ar) and Fig.~\ref{trbir}f (air) neglecting ionization but including higher-order Kerr coefficients \cite{loriot_measurement_2010}.
The shape of the signal clearly matches the plasma grating model better than the higher-order Kerr effect model.
The difference between Fig.~\ref{trbir}b and Fig.~\ref{trbir}c can be attributed mostly to the cumulative nature of the plasma nonlinearity, which also results in the slight asymmetry with respect to time in the spectrally integrated signal at high intensity with a transform-limited pulse observed in Fig.~\ref{loriot2}.

\subsection{Degenerate, chirped pulse spectral interferometry}
\begin{figure}
\begin{center}
\includegraphics[width=8.5cm]{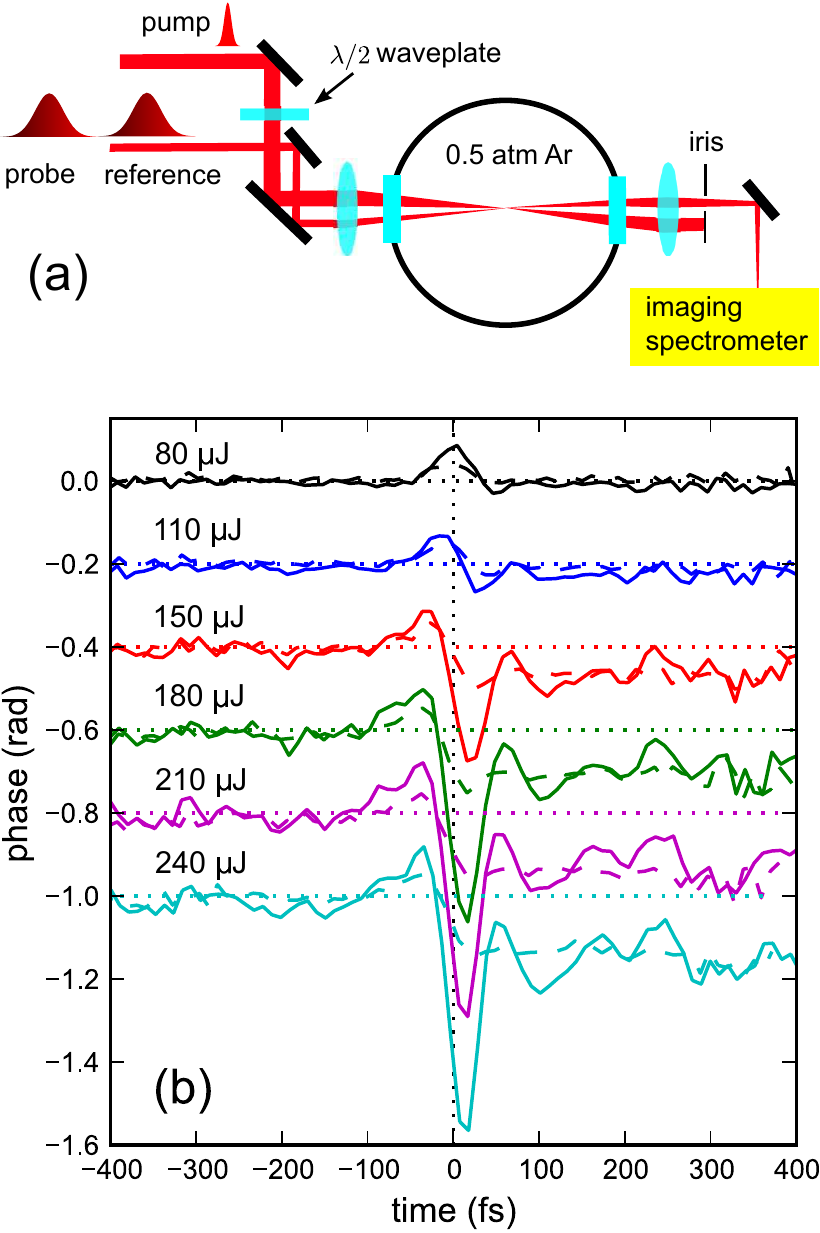}
\end{center}
\caption{Degenerate spectral interferometry results. (a) Experimental setup. (b) Extracted time domain phase shift as a function of the pump pulse energy for parallel polarizations (solid) and perpendicular (dashed).
}
\label{expt}
\end{figure}

To further investigate the origin of the negative birefringence, we use spectral interferometry with a nearly degenerate probe pulse.
The principle of the degenerate experiment is identical to our previous nondegenerate implementation using a supercontinuum probe \cite{chen_measurement_2007}.
The challenge in performing the experiment with a degenerate probe is rejecting pump light from the detection spectrometer.
We previously reported data using a degenerate probe with a polarization perpendicular to the pump, where polarizers were used to block the pump light before detection \cite{wahlstrand_measurements_2012}.
Here, we use a non-collinear geometry and reject pump light using an aperture, allowing orientation of the pump polarization parallel to the probe.
A sketch of the experimental apparatus is shown in Fig. \ref{expt}a.
The pump is a 40 fs FWHM pulse centered at 800 nm.
The pump and probe/reference beams cross at 3$^\circ$, allowing the pump beam to be blocked before probe detection.
The two beams are crossed inside a vacuum chamber backfilled with 0.5 atm of Ar.
The probe and reference pulses are chirped using a block of glass to a group delay dispersion of 1650 fs$^2$.

Results of the experiment are shown in Fig. \ref{expt}b. For the pump beam polarized perpendicular to the probe beam, the result is identical to that found with a collinear geometry \cite{wahlstrand_measurements_2012}.
We observe a positive instantaneous phase shift near the time of peak pump intensity ($t = 0$) due to the optical Kerr effect, and a negative, long-lasting signal from electrons freed by ionization.
When the pump beam is polarized parallel to the probe beam, we in addition observe a negative signal near $t = 0$ that is much larger than the positive Kerr signal at high intensity.
We attribute the oscillations in the data to interference between supercontinuum generated by the pump pulse and the probe pulse.
It is important to note that as the intensity is increased, the negative signal at $t=0$ appears simultaneously with the smooth plasma signal.
This is strong evidence for the plasma grating interpretation.
In addition, the shape of the signal is consistent with our calculation, shown in Fig.~\ref{sssi}.

\section{Conclusion}
We have presented a detailed theory and numerical calculations of pump-induced phase shifts of probe pulses in media with electronic, plasma, and rotational nonlinearities.
Such experiments are typically used to measure the nonlinear response of a material.
The theory includes the effect of the nonlinear interference grating formed by the space and time overlap of the pump and probe fields.
We showed that proper interpretation of the probe response, in terms of a medium's fundamental nonlinearities, can only be achieved by correctly accounting for the effect of the interference grating.
In particular, our calculations reveal how the presence of nonlinear interference gratings (in both the plasma and rotational responses) in recent pump-probe experiments have resulted in misinterpretation of the obtained results for the nonlinear response of gases at high intensity \cite{loriot_measurement_2009}.

In addition, we have presented results from two new degenerate pump-probe experiments, spectrally-resolved transient birefringence and noncollinear spectral interferometry, for the specific purpose of investigating the effect on a probe pulse of the nonlinear interference grating.
The results of both experiments are well reproduced by our calculations.
The results of this paper reinforce the idea that, for the purpose of measuring the nonlinear response of a medium using cross phase modulation, a nondegenerate pump-probe experiment is preferable to a degenerate one.
As shown earlier, for a nondegenerate experiment the only difference in the measured nonlinear refractive index and the nonlinear refractive index induced by a pulse and itself is a factor of 2 enhancement in the instantaneous, bound electronic nonlinearity.

\section*{Acknowledgments}
We acknowledge helpful discussions with Y.-H. Chen.
R.J.L. acknowledges support from the Office of Naval Research, the Air Force Office of Scientific Research, and the National Science Foundation.
H.M.M. acknowledges support from the Office of Naval Research, the National Science Foundation, the Department of Energy, and the Air Force Office of Scientific Research.

\appendix

\section{Electronic Kerr effect}
We put the total optical field $\bE = \bEp + \bEe$ into Eq.~(\ref{Pchi3}), recalling that the pump is polarized along $\bxhat$ and the probe is polarized in the $xy$ plane.
Keeping only terms containing $e^{i\bkp \cdot \br -i \omega_p t}$ and using our assumption $|A_e| \gg |A_p|$ to neglect terms second-order and higher in the probe field $A_p$, we find
\begin{eqnarray}
(\bPel)_x (\br, t) &=& \frac{2\pi}{n_0 c} \left[3\chit_{xxxx} A_x (\br, t) \right. \nonumber \\  &&\left. + \left(\chit_{xyxx}+\chit_{xxyx}+\chit_{xxxy} \right) A_y (\br, t) \right] \nonumber \\ && \times I_e (\br, t) e^{i \bkp \cdot \br-i\omega_p t}, \nonumber \\
(\bPel)_y (\br, t) &=& \frac{2\pi}{n_0 c} \left[3\chit_{yxxx} A_x(\br,t) \right. \nonumber \\ && \left. +   \left(\chit_{yyxx}+\chit_{yxyx}+\chit_{yxxy}\right) A_y(\br,t) \right] \nonumber \\ && \times I_e (\br, t) e^{i \bkp \cdot \br-i\omega_p t}, \nonumber
\end{eqnarray}
where we have rewritten the expressions in terms of the pump intensity.
In an isotropic medium, $\chit_{xyxx} = \chit_{xxyx} = \chit_{xxxy} =\chit_{yxxx} = 0$, so we neglect all terms containing those elements.
Taking the second derivative with respect to time, we find
\begin{equation*}
\frac{4\pi}{c^2} \frac{\partial^2(\bPel)_x}{\partial t^2} = \frac{2 n_0 n_2}{c^2} \frac{\partial^2}{\partial t^2} \left(I_e A_x e^{i \bkp \cdot \br-i\omega_p t} \right),
\end{equation*}
where $n_2$ is defined in the text.
Expanding the derivative, we find
\begin{multline*}
\frac{4\pi}{c^2} \frac{\partial^2(\bPel)_x}{\partial t^2} = \frac{2 n_0 n_2}{c^2} \left( \frac{\partial^2 I_e}{\partial t^2} A_x + 2 \frac{\partial I_e}{\partial t} \frac{\partial A_x}{\partial t} \right. \\ \left. -\omega_p^2 I_e A_x -2i\omega_p \frac{\partial I_e}{\partial t} A_x -2i \omega_p I_e \frac{\partial A_x}{\partial t} +\frac{\partial^2 A_x}{\partial t^2} \right) \\ \times e^{i \bkp \cdot \br-i\omega_p t}.
\end{multline*}
For many-cycle pulses, only a few terms are important above.
In the SVEA, second-order and higher time derivatives are neglected.
This leads to Eq.~(\ref{finalKerr}).
In an isotropic medium far away from any resonances, $\chi^{(3)}_{yyxx} = \chi^{(3)}_{yxyx} = \chi^{(3)}_{yxxy} = \chi^{(3)}_{xxxx}/3$, from which one can similarly derive Eq.~(\ref{finalKerr2}).

\begin{widetext}
\section{Rotational grating}
The quantity $\langle j,m | \sin 2\theta \cos \phi | j',m' \rangle$ can be expressed in terms of spherical harmonics,
\begin{eqnarray}
\langle j,m | \sin 2\theta \cos \phi | j',m' \rangle &=& \langle j,m | \sin \theta \cos \theta (e^{i\phi}+e^{-i\phi}) | j',m' \rangle \nonumber \\ 
&=& \sqrt{\frac{8\pi}{15}} \left( \langle j,m|Y_2^1(\theta,\phi) |j',m' \rangle -  \langle j,m| Y_2^{-1} (\theta,\phi) |j',m' \rangle \right). \nonumber
\end{eqnarray}
These integrals of 3 spherical harmonics lead to \cite{abramowitz}
\begin{multline}
\langle j,m | \sin 2\theta \cos \phi | j',m' \rangle = \left\{ \left[ A^{m}_+ (j') \delta_{m,m'-1} -A_+^{-m} (j') \delta_{m,m'+1} \right] \delta_{j,j'+2} \right. \\ \left. + \left[ A^{m}_0 (j') \delta_{m,m'-1} -A_0^{-m} (j') \delta_{m,m'+1} \right] \delta_{j,j'} + \left[ A^{m}_- (j') \delta_{m,m'-1} -A_-^{-m} (j') \delta_{m,m'+1} \right] \delta_{j,j'-2} \right\},
\label{sin2thetacosphi}
\end{multline}
where
\begin{subequations}
\begin{eqnarray}
A_+^m (j) &=& \left[ \frac{(j-m)(j-m+1)[(j+2)^2-m^2]}{(2j+1)(2j+3)^2(2j+5)} \right]^{1/2}, \\
A_0^m (j) &=& -\frac{(1+2m)[(j-m)(j+1+m)]^{1/2}}{(2j-1)(2j+3)}, \\
A_-^m (j) &=& -\left[ \frac{(j+m+1)(j+m)[(j-1)^2-m^2]}{(2j-3)(2j-1)^2(2j+1)} \right]^{1/2}.
\end{eqnarray}
\label{Acoeffs}
\end{subequations}
Similarly,
\begin{equation}
\langle j,m | \cos^2\theta | j',m' \rangle = B_+^m(j')\delta_{j,j'+2} \delta_{mm'} + B_-^m(j')\delta_{j,j'-2}\delta_{mm'}+\left[\frac{1}{3} + B_0^m(j') \right] \delta_{jj'} \delta_{mm'},
\label{cos2theta}
\end{equation}
where
\begin{subequations}
\begin{eqnarray}
B_+^m (j) &=& \left[ \frac{[(j+2)^2-m^2][(j+1)^2-m^2]}{(2j+1)(2j+3)^2(2j+5)} \right]^{1/2}, \\
B_0^m (j) &=& \frac{2}{3} \frac{j(j+1)-3m^2}{(2j-1)(2j+3)}, \\
B_-^m (j) &=& \left[ \frac{[(j^2-m^2)[(j-1)^2-m^2]}{(2j-1)(2j+1)^2(2j-3)} \right]^{1/2}.
\end{eqnarray}
\end{subequations}

Using Eqs.~(\ref{sin2thetacosphi},\ref{cos2theta}) in Eqs.~(\ref{Hxx},\ref{Hxy}) and those plus Eq.~(\ref{rho0}) in Eq.~(\ref{rho1}), we find that only certain elements of $\rho^{(1)}$ are nonzero.
These are
\begin{subequations}
\begin{eqnarray}
\rho^{(1)}_{(j,m),(j-2,m)} (\br,t) &=& -B_- (j-2)\left(\rho^{(0)}_j-\rho^{(0)}_{j-2} \right) F_{j,j-2} (\br,t)\\
\rho^{(1)}_{(j,m),(j+2,m)} (\br,t) &=& -B_+ (j+2) \left(\rho^{(0)}_j-\rho^{(0)}_{j+2} \right) F_{j,j+2} (\br,t)\\
\rho^{(1)}_{(j,m),(j-2,m+1)} (\br,t) &=& - A_+^m (j-2) \left(\rho^{(0)}_j-\rho^{(0)}_{j-2} \right) G_{j,j-2} (\br,t) \\
\rho^{(1)}_{(j,m),(j-2,m-1)} (\br,t) &=&  A_+^{-m} (j-2)\left(\rho^{(0)}_j-\rho^{(0)}_{j-2} \right) G_{j,j-2} (\br,t) \\
\rho^{(1)}_{(j,m),(j+2,m+1)} (\br,t) &=& - A_-^m (j+2) \left(\rho^{(0)}_j-\rho^{(0)}_{j+2} \right) G_{j,j+2} (\br,t) \\
\rho^{(1)}_{(j,m),(j+2,m-1)} (\br,t) &=&  A_-^{-m} (j+2) \left(\rho^{(0)}_j-\rho^{(0)}_{j+2} \right) G_{j,j+2} (\br,t),
\end{eqnarray}
\label{rho1e}
\end{subequations}
where
\begin{eqnarray}
F_{jj'} (\br, t) &=& \frac{i\Delta \alpha}{2\hbar}\int_{-\infty}^t e^{i\omega_{j,j'}(t'-t)} [|A_e|^2 (\br,t') +(A_e^* (\br, t') A_x (\br, t') e^{i\Delta \bk \cdot \br-i\Delta \omega t'}+c.c.)] dt'. \nonumber \\
G_{jj'} (\br, t) &=& \frac{i\Delta \alpha}{4\hbar}\int_{-\infty}^t e^{i\omega_{j,j'}(t'-t)} (A_e^* (\br, t') A_y (\br, t') e^{i\Delta \bk \cdot \br-i\Delta \omega t'} +c.c.)dt'. \nonumber
\end{eqnarray}

Finally,
\begin{eqnarray}
\langle \sin 2\theta \cos \phi \rangle_t &=& \mathrm{Tr} (\bm{\rho} \sin 2\theta \cos \phi) = \sum \rho^{(1)}_{(j,m),(j',m')}(\br, t) \langle j,m| \sin 2\theta \cos \phi |j',m' \rangle \nonumber \\
&=& \sum \left[ -A_-^{-m-1} (j) \rho^{(1)}_{(j,m),(j-2,m+1)}(\br, t) +A_-^{m-1} (j) \rho^{(1)}_{(j,m),(j-2,m-1)}(\br, t)\right. \nonumber\\  && \left. - A_+^{-m-1} (j) \rho^{(1)}_{(j,m),(j+2,m+1)}(\br, t)+A_+^{m-1} (j) \rho^{(1)}_{(j,m),(j+2,m-1)} (\br, t)\right],\nonumber \\
&=& \sum \left[ -A_-^{-m-1} (j) \rho^{(1)}_{(j,m),(j-2,m+1)}(\br, t) +A_-^{m-1} (j) \rho^{(1)}_{(j,m),(j-2,m-1)}(\br, t)\right. \nonumber\\  && \left. - A_+^{-m} (j) \rho^{(1)}_{(j,m-1),(j+2,m)}(\br, t)+A_+^{m} (j) \rho^{(1)}_{(j,m+1),(j+2,m)}(\br, t) \right].\nonumber
\end{eqnarray}
Inserting Eqs.~(\ref{rho1e}), we find
\begin{eqnarray}
\langle \sin 2\theta \cos \phi \rangle_t &=& \sum_{l,m} \left[ A_-^{-m-1} (j) A_+^m (j-2) \left(\rho^{(0)}_j-\rho^{(0)}_{j-2} \right) G_{j,j-2} (\br,t) \right. \nonumber\\
&& + A_-^{m-1} (j) A_+^{-m} (j-2)\left(\rho^{(0)}_j-\rho^{(0)}_{j-2} \right) G_{j,j-2} (\br,t) \nonumber\\
&& - A_+^{-m}(j-2) A_-^m (j+2) \left(\rho^{(0)}_j-\rho^{(0)}_{j+2} \right) G_{j+2,j} (\br,t) \nonumber\\
&& \left. - A_+^m(j-2) A_-^{-m} (j+2) \left(\rho^{(0)}_j-\rho^{(0)}_{j+2} \right) G_{j+2,j} (\br,t) \right] \nonumber\\
&=& - \sum_{j,m} \left[ A_-^{-m-1} (j) A_+^m(j-2) + A_-^{m-1} (j) A_+^{-m} (j-2) \right] (\rho^{(0)}_j - \rho^{(0)}_{j-2}) H_{j,j-2} (\br,t) \label{finalsin2thetacosphi}
\end{eqnarray}
where in the last line we have used
\begin{equation*}
G_{jj'} (\br,t) - G_{j'j} (\br,t) = H_{jj'} (\br,t) \equiv \frac{\Delta\alpha}{2\hbar} \int_{-\infty}^t \sin [\omega_{j,j'}(t'-t)] [A_e^* (\br,t) A_y (\br,t)e^{i\Delta \bk \cdot \br-i\Delta \omega t'}+c.c.] dt'.
\end{equation*}
Using Eqs.~(\ref{Acoeffs}) in Eq.~(\ref{finalsin2thetacosphi}) and cancelling many factors leads to Eq.~(\ref{sin2thetacosphit}).
A similar calculation, which we do not reproduce here because it has been described previously \cite{chen_single-shot_2007}, leads to Eqs.~(\ref{cos2thetast},\ref{cos2thetagt}).
\end{widetext}

\end{document}